\documentclass[slac_one]{revtex4}
\usepackage{graphicx}
\usepackage{fancyhdr}
\usepackage{subfigure} 
\def\missET {{\not\!\! E_T}}

\begin{document}

\title{Top Quark Mass in the Dilepton and Alljets Channels}%% Paper title goes here

\author{Bodhitha Jayatilaka (on behalf of the CDF and D0 Collaborations)}
\affiliation{Duke University, Durham, NC 27708, USA}

\begin{abstract}
Recent measurements of the top quark mass in the dilepton ($t\bar{t}\to \bar{b}\ell^{-}\bar{\nu_{\ell}}b \ell^{'+}\nu^{'}_{\ell}$) and alljets ($t\bar{t}\to b\bar{b}q\bar{q}'q'\bar{q}$) decay channels from the CDF and D0 collaborations are presented. These decay channels provide unique challenges from the lepton+jets decay channel. Results utilizing up to 2.8 fb$^{-1}$ of $p\bar{p}$ collider data from the Fermilab Tevatron are discussed.
\end{abstract}
\maketitle

\thispagestyle{fancy}

\section{INTRODUCTION} % Section title should be in all capitals.

Precision measurement of the top quark mass, $M_t$, places constraints on the masses to which the top quark contributes radiative corrections, including the yet unobserved Higgs boson, as well as helps place constraints on the masses of particles in extensions to the standard model.

 At present, the Fermilab Tevatron, a $p\bar{p}$ collider with $\sqrt{s}=1.96$~TeV, is the only accelerator capable of producing top quarks for study, where they are primarily produced in pairs and decay nearly 100\% of the time as $t\bar{t}\to W^{+}bW^{-}\bar{b}$. The decay of the two $W$ bosons determines the decay channel of the $t\bar{t}$ system. While the most precise measurements of $M_t$ are made using the lepton+jets channel ($t\bar{t}\to \bar{b}\ell\nu_{\ell}b q\bar{q}'$) significant contributions to the overall knowledge of $M_t$ are made with measurements in the dilepton ($t\bar{t}\to \bar{b}\ell^{-}\bar{\nu_{\ell}}b \ell^{'+}\nu^{'}_{\ell}$) and alljets ($t\bar{t}\to b\bar{b}q\bar{q}'q'\bar{q}$) channels. In addition, these decay channels offer unique experimental challenges not present in the lepton+jets channel. 
 
 We discuss several recent measurements of the top quark mass in the dilepton channel from both CDF and D0 as well as measurements in the alljets channel from CDF. These measurements use 2.0-2.8 fb$^{-1}$ of Run II collider data. 
 
\section{MEASUREMENT STRATEGIES}

Two broad strategies are utilized in measuring $M_t$. Matrix element (ME) techniques use leading-order matrix element calculations convoluted with detector resolution functions to estimate per-event probabilities for an assumed $M_t$. Similar probabilities can be calculated for dominant background processes to reduce the impact of background-like events in the final sample. These per-event probabilities for the entire data sample are then combined to extract the most probable value of $M_t$. CDF and D0 both utilize ME methods to measure $M_t$ in the dilepton channel. Template-based methods reconstruct a single mass for each event. Templates of reconstructed mass are then formed using Monte Carlo (MC) simulated events in both signal and background. The distribution of reconstructed masses in the data sample is then compared to these templates to extract a most probable value of $M_t$. CDF and D0 both utilize template-based methods to measure $M_t$ in the dilepton mass. CDF  also uses a template-based method to measure $M_t$ in the alljets channel. Additionally, CDF uses a method that combines features of both ME and template-based analyses, known as the ideogram method, to measure $M_t$ in the alljets channel. 

\section{DILEPTON DECAY CHANNEL}

\subsection{Decay channel and event selection}

The dilepton decay channel consists of $t\bar{t}$ events where both $W$ bosons decay to a charged lepton and neutrino. This decay channel has the smallest branching ratio of $t\bar{t}$ decay channels, approximately 5\% The final state in dilepton events consists of two charged leptons ($e$ or $\mu$), two energetic jets resulting from the $b$ quarks, and large missing transverse energy ($\missET$) from the two neutrinos. As there are two neutrinos and only one measurement of $\missET$, the dilepton channel is kinematically under-constrained and at least one variable must be integrated over in order to reconstruct an event mass. The dilepton channel is, however, largely free of  QCD background that is present in other $t\bar{t}$ decay channels. The remaining background largely consists of Drell-Yan ($Z/\gamma^*\to e^+e^-,\mu^+\mu^-$), diboson ($WW, WZ, ZZ$) and instrumental background where an object is incorrectly reconstructed as a lepton. 

Both CDF and D0 have employed cut-based event selections for measurements in the dilepton decay channel. Two broad strategies employed for cut-based selections are using two well identified leptons in the final state, or looking for one identified lepton and one isolated track. The former yields better signal purity but lower statistics while the latter allows more statistics at the expense of more background. A third strategy employed by CDF is to optimize an event selection specifically for precision in $M_t$ measurement. This is accomplished by the usage of neuroevolution. Beginning with a population of random neural networks, $M_t$ measurement precision is evaluated for each network. Poor performers are culled and strong performers are bred together and mutated in successive generations until performance reaches a plateau. The resulting best-performing network is used as the basis of the final event selection. This approach results in an increase in signal yield at the expense of a large increase in well-identifiable backgrounds. Event yields for examples of all 3 selection approaches are provided in Table~\ref{tab:selection}.

\begin{table}
\caption{Event yields for lepton+isolated track event selection, two identified leptons, and evolutionary neural network. The D0 selection shown is only in the $e\mu$ final state. Additionally, D0 also utilizes a lepton+isolated track selection (not shown in table).}
\begin{tabular}{lccc}
Source & CDF (lepton+track) & D0 (two lepton) & CDF (neuroevolution) \\
\hline
$t\bar{t}$ ($M_t = 175$~GeV$/c^2$)& $164.3\pm 5.1$ & $89.6\pm 3.8$ & $121.8\pm 7.5$\\
Drell-Yan & $50.1\pm 7.2$ & $12.8\pm 1.4$ & $140.2\pm 14.5$\\
Mis-ID leptons & $81.0\pm 15.9$ & $7.1\pm 1.2$ & $33.5\pm 5.9$ \\
Diboson & $15.4\pm 1.0$ & $4.0\pm 0.6$ & $18.0\pm 3.7$ \\ \hline
Total expected & $311.8\pm 18.7$ & $110.8\pm 5.0$ & $313.3\pm 21.2$ \\ \hline
Data ($\int\mathcal{L} dt$) & 330 (2.8 fb$^{-1}$) & 107 (2.8 fb$^{-1}$) & 344 (2.0 fb$^{-1}$)\\
\end{tabular}
\label{tab:selection}
\end{table}

\subsection{Measurements}
\begin{figure}
 \subfigure[]{
 \includegraphics[width=0.24\columnwidth]{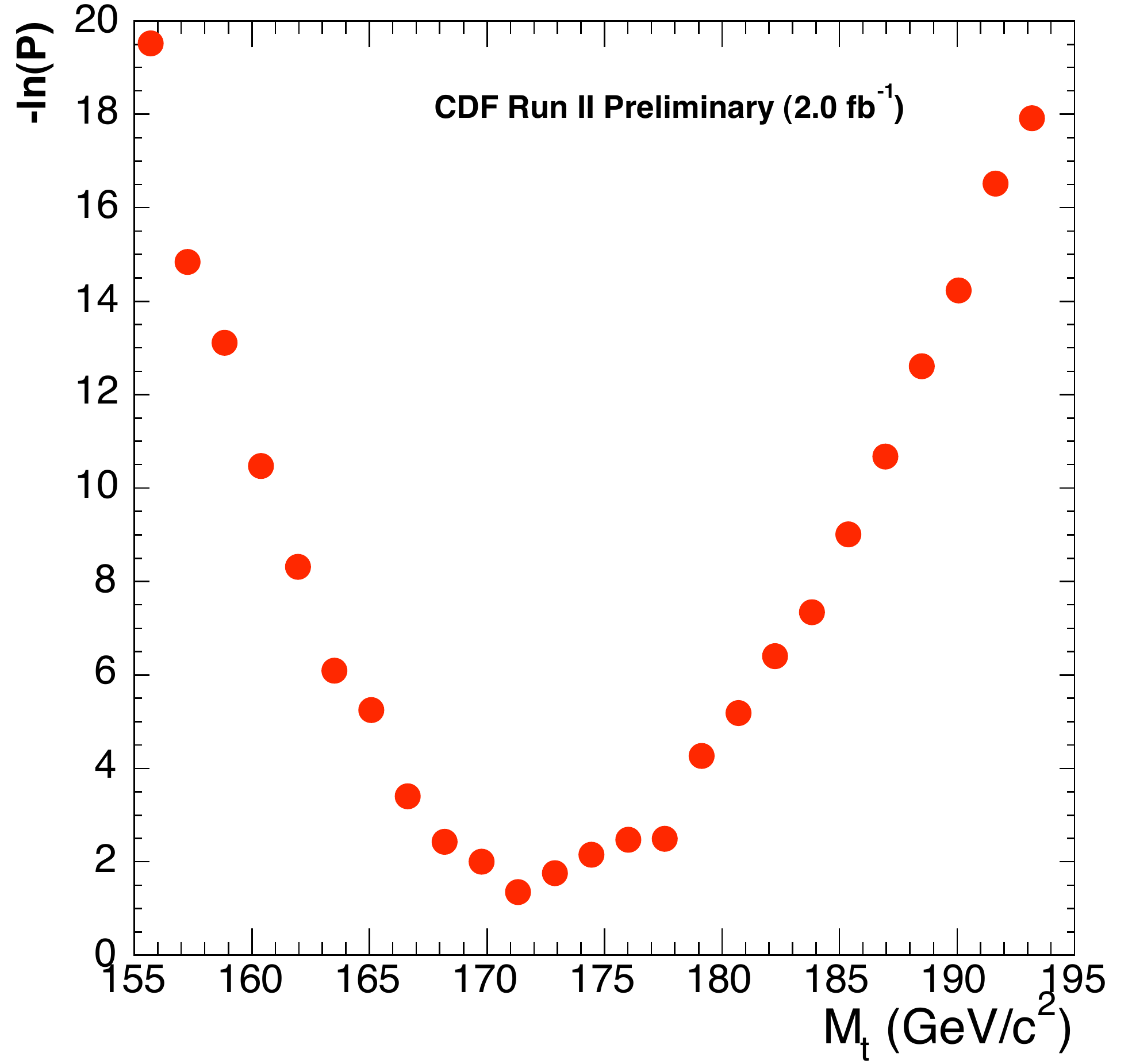}
  \label{fig:cdfme}
  }
   \subfigure[]{
 \includegraphics[width=0.24\columnwidth]{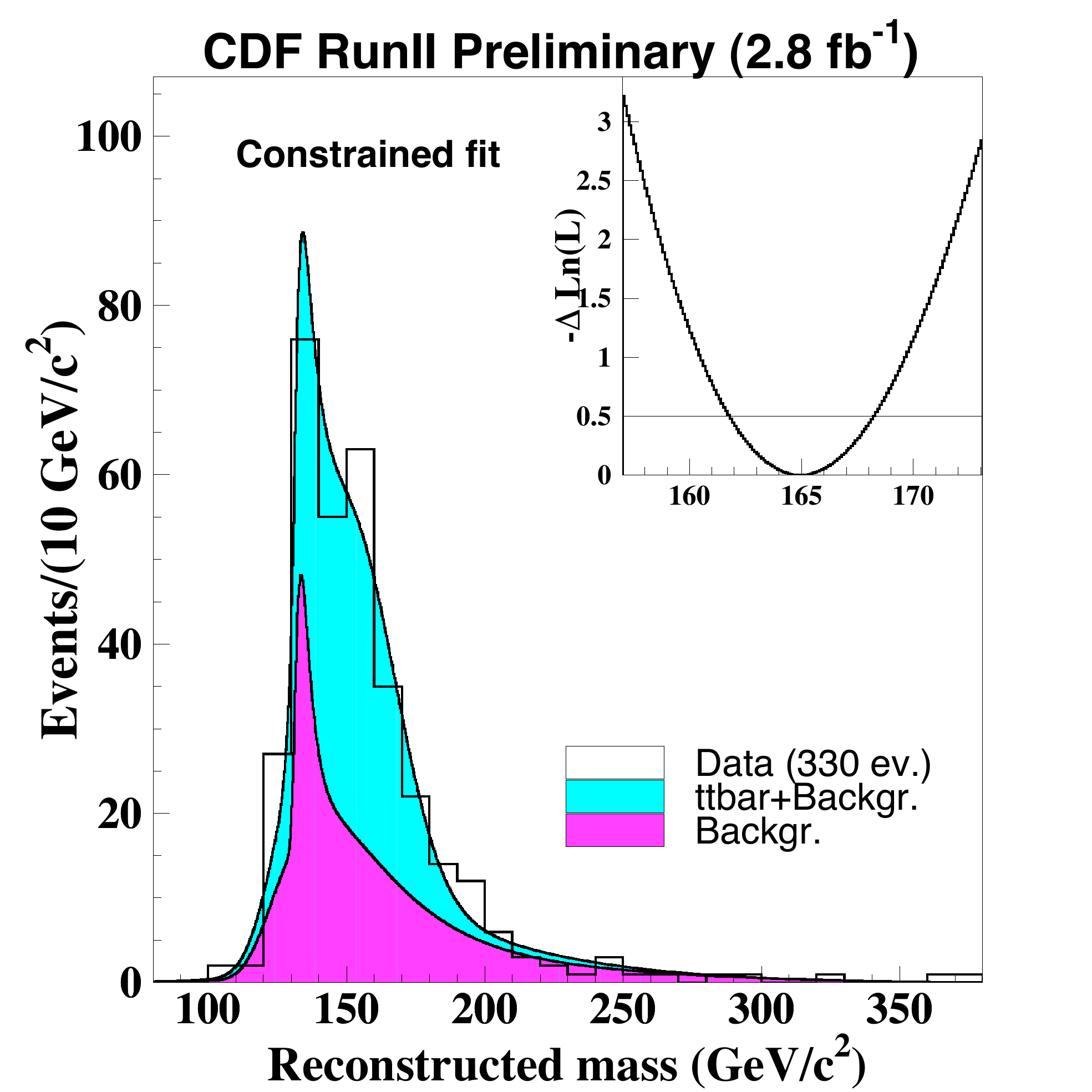}
  \label{fig:cdftempdil}
  }
   \subfigure[]{
 \includegraphics[width=0.24\columnwidth]{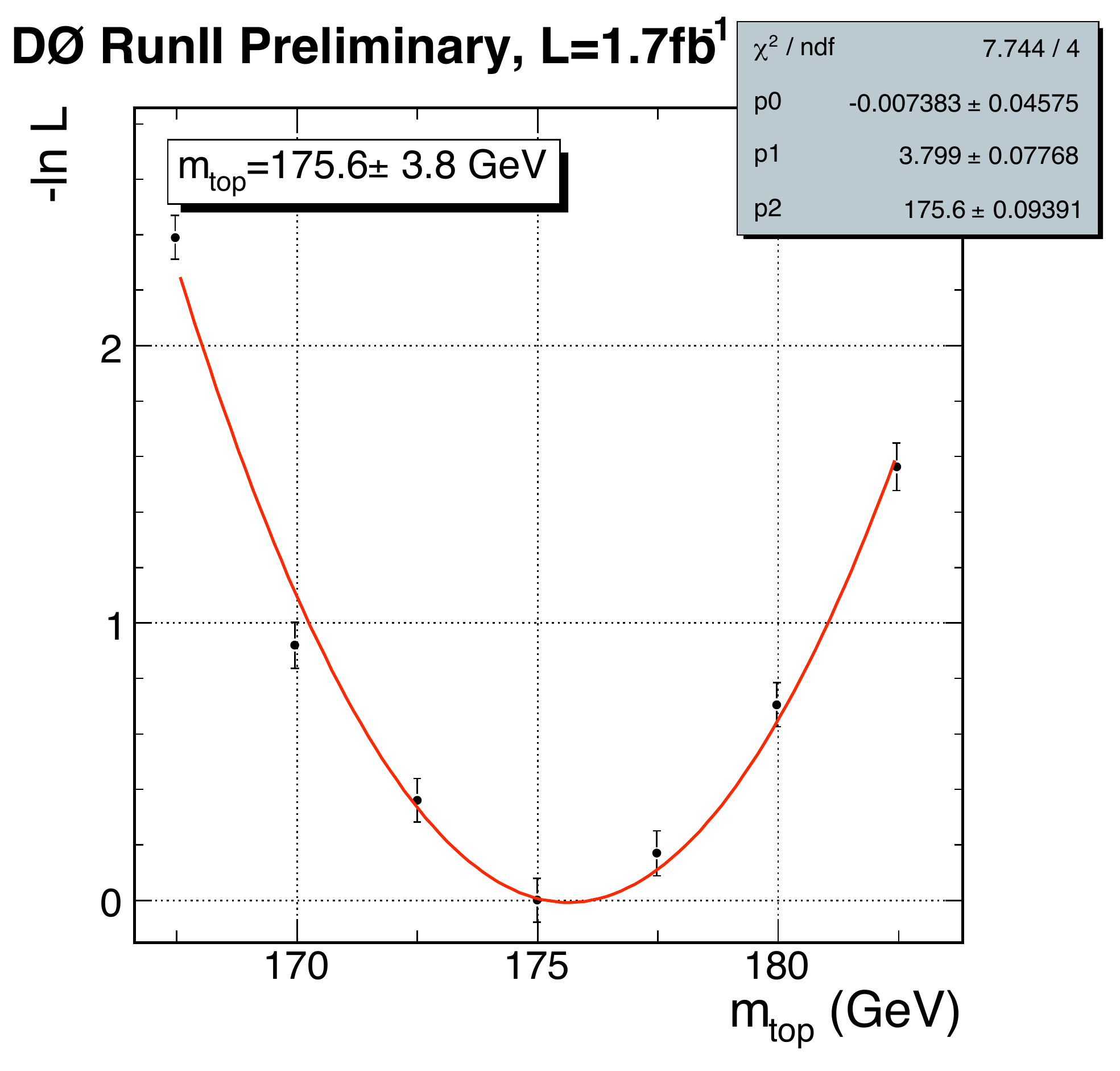}
  \label{fig:cdftempdil}
  }
  \caption{(a) Negative log of probability for the data sample used in the CDF matrix element-based measurement in the dilepton channel. (b) Reconstructed mass distribution for the data sample used in the CDF template-based measurement in the dilepton channel. The inset shows the negative log likelihood used in the final fit. (c) Negative log likelihood used in the fit for the D0 matrix element-based measurement in the dilepton channel. This plot is for the 1.7 fb$^{-1}$ from Run IIb. The final result uses a combined fit from Run IIa and Run IIb.}
  \label{fig:dilresult}
 \end{figure}

Both CDF and D0 perform measurements of $M_t$ using matrix element and template-based methods in the dilepton channel. CDF performs a matrix element-based measurement using a neuroevolution selected sample of events in 2.0 fb$^{-1}$ of data~\cite{dilmecdf}. The resulting measurement is $M_t = 171.2\pm 2.7 (\mathrm{stat.})\pm 2.9 (\mathrm{syst.})$~GeV$/c^2$. The probability distribution for this result is shown in Fig.~\ref{fig:dilresult}(a). D0 performs a matrix element-based measurement using a two-identified lepton selection in 2.8~fb$^{-1}$ of data~\cite{dilmed0}. This measurement is performed in the $e^{\pm}\mu^{\mp}$ final state and results in $M_t= 172.9\pm 3.6 (\mathrm{stat.})\pm 2.3 (\mathrm{syst.})$~GeV$/c^2$. The resulting likelihood distribution for this result is shown in Fig.~\ref{fig:dilresult}(c). 

CDF performs a template-based measurement integrating over neutrino $\phi$ using a lepton+isolated track sample in 2.8 fb$^{-1}$ of data~\cite{diltempcdf}. This measurement yields $M_t = 165.1^{+3.3}_{-3.1} (\mathrm{stat.})\pm 3.1 (\mathrm{syst.})$~GeV$/c^2$. The reconstructed mass distribution and likelihood distribution for this analysis are shown in Fig.~\ref{fig:dilresult}(b). D0 has recently performed a template-based measurement integrating over neutrino $\eta$ using a lepton+isolated track sample~\cite{diltempd0}. This measurement yields $M_t = 176.0\pm 5.3 (\mathrm{stat.})\pm 2.0 (\mathrm{syst.})$~GeV$/c^2$. Additionally, D0 has performed a statistical combination of $M_t$ measurements in the dilepton channel~\cite{dilcombd0}. This combination yields $M_t = 174.4\pm 3.2 (\mathrm{stat.})\pm 2.1 (\mathrm{syst.})$~GeV$/c^2$.

A list of systematic uncertainties for the measurements discussed here is shown in Table~\ref{tab:dilsyst}. In all analyses presented, the systematic uncertainties are dominated by the uncertainty in the jet energy scale (JES). While an {\it in situ} measurement of the JES is possible in the alljets and lepton+jets decay channels, the absence of a hadronically decaying $W$ boson makes this impossible in the dilepton channel.

\section{ALLJETS DECAY CHANNEL}

\subsection{Decay channel and event selection}

The alljets decay channel consists of $t\bar{t}$ events where both $W$ bosons decay to quarks. This decay has the largest branching ratio of all $t\bar{t}$ decay channels, approximately 44\%, at the expense of a very large QCD background. The final state in alljets events consist of six or more energetic jets, two resulting from $b$ quarks from top decay and four from quarks from the decay of the $W$s. While this is a kinematically fully reconstructable state, a large number of permutations of jet-quark assignments that must be considered when reconstructing the mass.
 
 The presence of a large multi-jet background makes identification of one or more $b$ jet via secondary vertex tagging, or $b$-tagging as well as the application of a neural network (NN) based selection a necessity~\cite{allhadprd}. In the template based analysis, events with 1 or 2 $b$-tags and $6\geq N_{jets}\geq 8$ are selected following a NN cut. In 1-tag (2-tag) events resulting from this selection, $2409\pm 68$ ($338\pm 28$) background events are expected, while 2881 (537) events are observed in 2.1~fb$^{-1}$ of data. In the ideogram-based analysis, events with $\geq 2$ $b$-tags and exactly 6 jets are selected after a NN cut. Here, 301 events are observed in 1.9~fb$^{-1}$ of data, of which 129 events are expected to be $t\bar{t}$.

% In the template-based analysis, a NN is trained to further separate $t\bar{t}$ events from QCD background events after secondary vertex tagging is required~\cite{allhadprd}. This NN is evaluated separately for events with one $b$-tag and events with two $b$-tags. After NN selection, in 1-tag (2-tag) events, $2409\pm 68$ ($338\pm 28$) background events are expected, while 2881 (537) events are observed in 2.1~fb$^{-1}$ of data. In the ideogram-based analysis, a NN is trained to separate background prior to the requirement of $b$-tags. After the tagging requirement is enforced, 301 events are observed in 1.9~fb$^{-1}$ of data, of which 129 events are expected to be $t\bar{t}$.
 
\subsection{Measurements}

\begin{figure}
  \subfigure[]{
 \includegraphics[width=0.23\columnwidth]{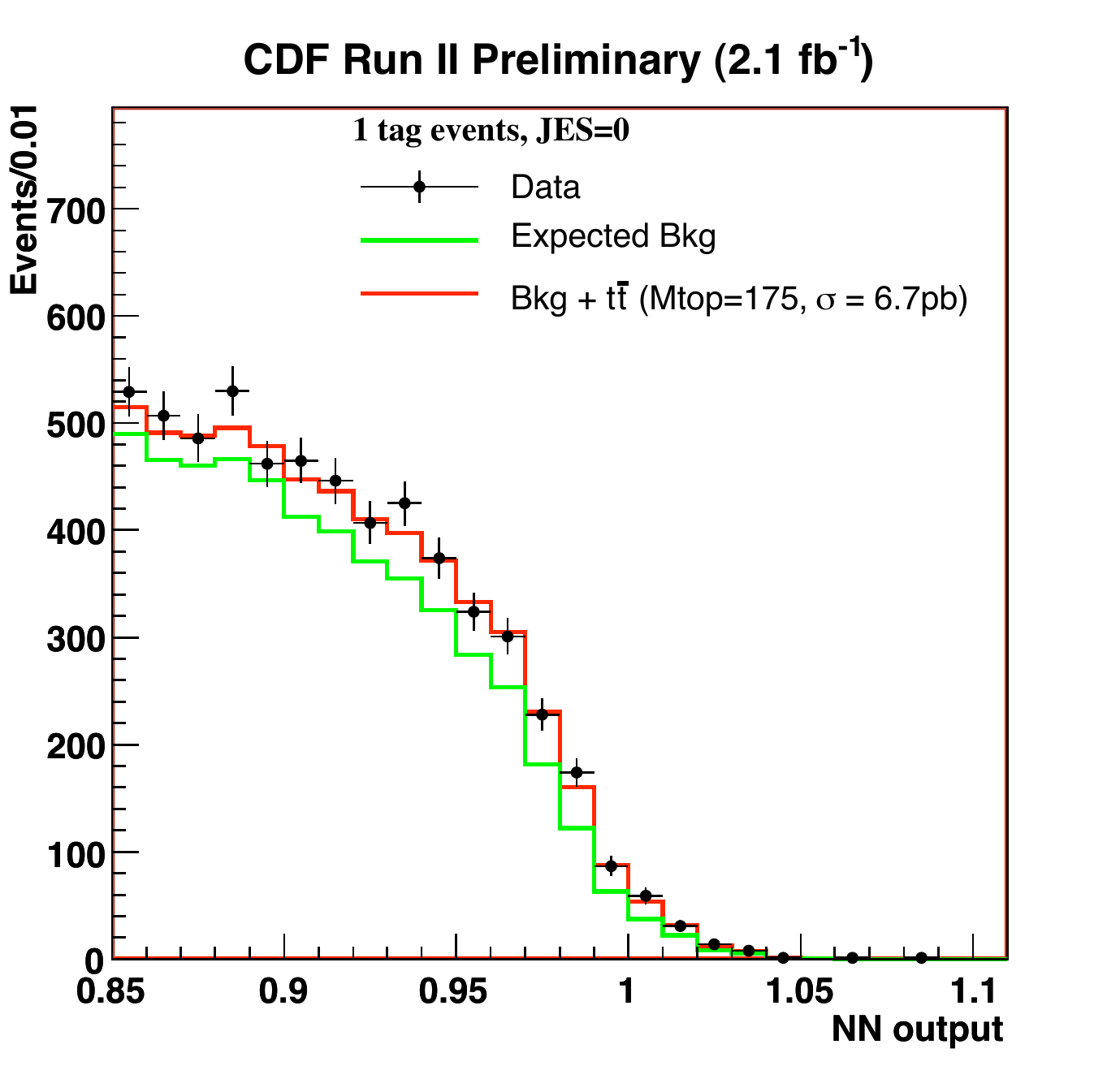}
  \label{fig:cdfme}
  }
 \subfigure[]{
 \includegraphics[width=0.25\columnwidth]{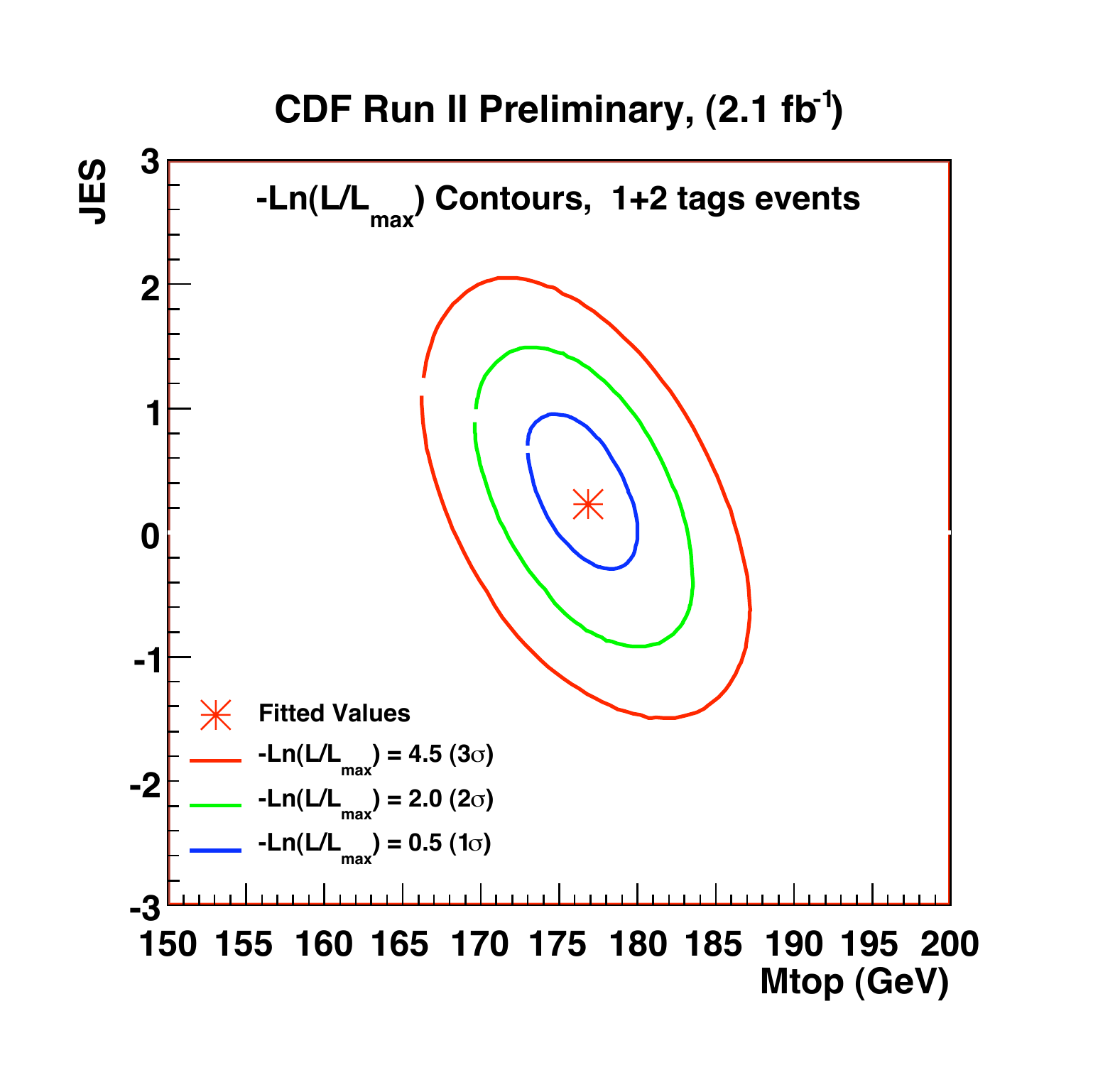}
  \label{fig:cdfme}
  }
   \subfigure[]{
 \includegraphics[width=0.33\columnwidth]{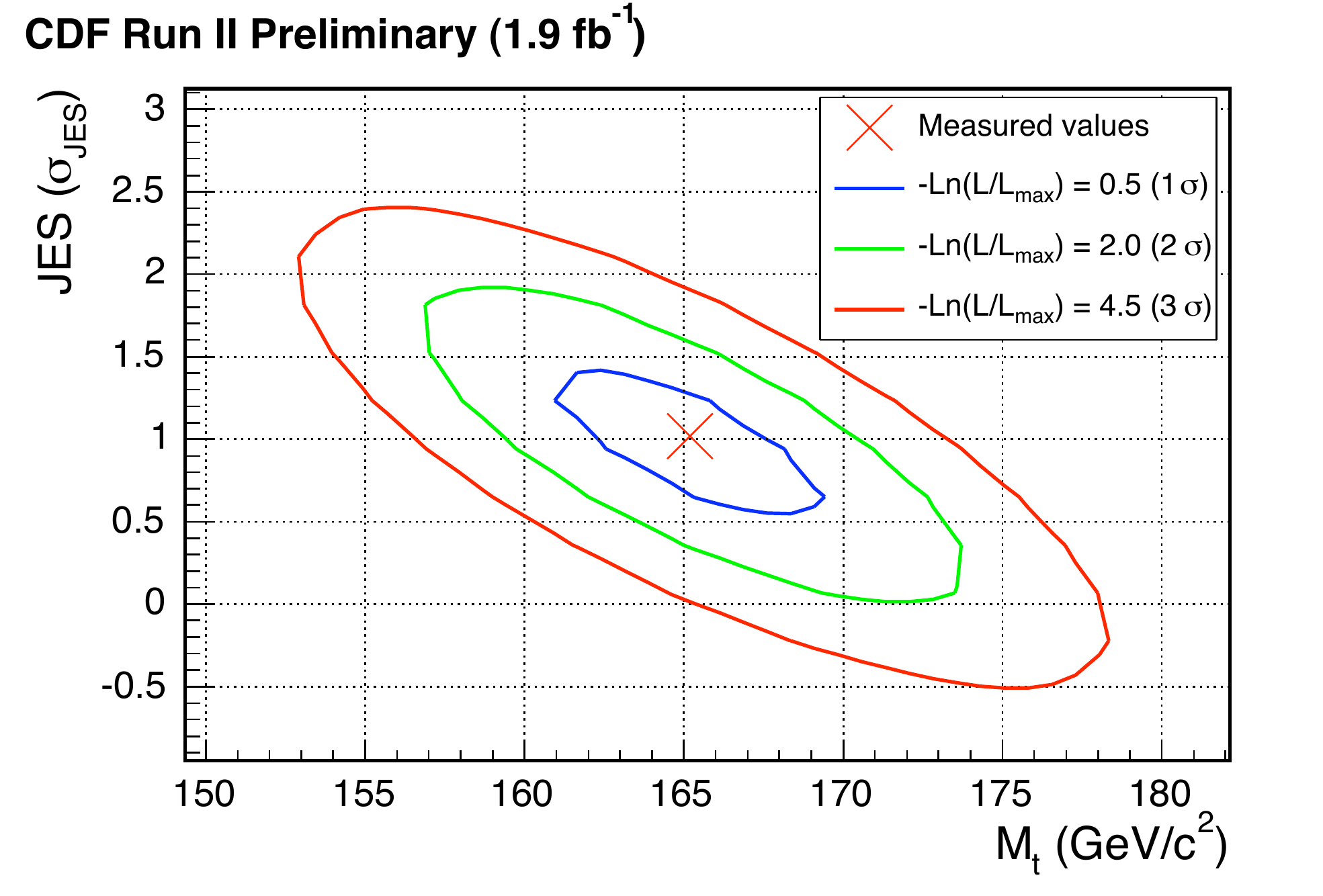}
  \label{fig:cdftempdil}
  }
    \caption{(a) NN output for alljets selection for events with one $b$-tag. Two-dimensional likelihood plot for the (b) template-based and (c) ideogram-based measurements of $M_t$ in the alljets channel.}
  \label{fig:alljetsresult}
 \end{figure}

CDF performs a template-based measurement using 2.1 fb$^{-1}$ of data. This measurement utilizes a two-parameter fit to extract both $M_t$ and the JES~\cite{allhadtemp}. The resulting likelihood is shown in Fig.~\ref{fig:alljetsresult}(a) is used to measure $M_t = 176.9\pm 3.8(\mathrm{stat.+JES})\pm 1.7(\mathrm{syst.})$~GeV$/c^2$. CDF also performs an ideogram-based measurement using 1.9~fb$^{-1}$ of data also making an {\it in situ} measurement of the JES~\cite{allhadideo}. The two dimensional likelihood is shown in Fig.~\ref{fig:alljetsresult}(b). This analysis yields $M_t = 165.2\pm 4.4(\mathrm{stat.+JES})\pm 1.9(\mathrm{syst.})$~GeV$/c^2$.

 A list of systematic uncertainties apart from the {\it in situ} JES for these two measurements is provided in Table~\ref{tab:dilsyst}.

\section{CONCLUSION}

CDF and D0 have both completed measurements of $M_t$ in the dilepton channel using a variety of techniques. CDF has also completed measurements of $M_t$ in the alljets channel. Measurements from these two channels contribute approximately 12\% of the weight in the most recent Tevatron combination of $M_t$ measurements~\cite{tevmtop}.

\begin{table}
\caption{Summary of systematic uncertainties for $M_t$ measurements in the dilepton and alljets decay channel. All uncertainties are in units of GeV$/c^2$. Residual JES refers to any uncertainty in the JES not accounted for by {\it in situ} measurement of it.}
\begin{tabular}{l|cccc|cc}
& \multicolumn{4}{c|}{Dilepton channel}&\multicolumn{2}{c}{Alljets channel}\\
Source & CDF ME  & CDF Template  & D0 ME  & D0 Template & CDF Template & CDF Ideogram \\
\hline
Residual JES & 2.5 & 2.9 & 2.3 & 1.7 & 0.8 & 0.7\\
$b$ specific JES & 0.4 & 0.4 & 0.3 & 0.5 & 0.6 & 0.3\\
Jet resolution & -- & -- & 0.7 & -- & -- & --\\
Pileup & 0.2 & 0.2 & -- & -- & 0.3 & 0.7\\
Monte Carlo Statistics & 0.5 & -- & -- & 0.1 & 1.1 & 0.1 \\
Parton Distr. Functions & 0.6 & 0.3 & 0.2 & 0.3 & 0.4 & 0.4\\
Generator & 0.9 & 0.2 & -- & 0.8 & 0.5 & 0.8\\
Background Shape & 0.2 & 0.5 & -- & 0.3 & -- & 0.4\\
QCD Radiation & 0.5 & 0.3 & 0.4 & 0.1 & 0.5 & 1.2\\
Sample Composition & 0.3 & 0.5 & 0.3 & 0.3 & 0.5 & 0.5\\
Lepton Resolution & 0.1 & 0.3 & 0.3 & 0.1 & N/A & N/A\\
\hline
Total & 2.9 & 3.1 & 2.5 & 2.0 & 1.7 & 1.9\\
\end{tabular}
\label{tab:dilsyst}
\end{table}

\begin{acknowledgments}
The author acknowledges the CDF and D0 collaborations for their hard work in producing the results presented here and the U.S. Department of Energy for support.
\end{acknowledgments}

\end{document}